\documentclass[journal=jacsat,manuscript=article]{achemso}
\usepackage[version=3]{mhchem}
\usepackage{xcolor}
\usepackage{graphicx}
\usepackage{dcolumn}
\usepackage{bm}
\usepackage{hyperref}%

\author{Vikash Sharma}
\email{vikash.sharma@tifr.res.in}
\affiliation{UGC-DAE Consortium for Scientific Research
	University campus, Khandwa road, Indore-452001, Madhya Pradesh, India}
\altaffiliation{Presently at Department of Condensed Matter Physics and Materials Science, Tata Institute of Fundamental Research\\ Homi Bhabha Road, Colaba, Mumbai-400005, India.}
\author{Sudip Pal}
\affiliation{UGC-DAE Consortium for Scientific Research
	University campus, Khandwa road, Indore-452001, Madhya Pradesh, India}
\author{Divya Sharma}
\affiliation{Govt. Girls PG College, Ujjain-456010, MP,	India}
\author{Dinesh Kumar Shukla}
\affiliation{UGC-DAE Consortium for Scientific Research
	University campus, Khandwa road, Indore-452001, Madhya Pradesh, India}
\author{Ram Janay Chaudhary}
\affiliation{UGC-DAE Consortium for Scientific Research
	University campus, Khandwa road, Indore-452001, Madhya Pradesh, India}
\author{Gunadhor Singh Okram}
\affiliation{UGC-DAE Consortium for Scientific Research
	University campus, Khandwa road, Indore-452001, Madhya Pradesh, India}
\title{Size-induced Exchange Bias in Single-phase CoO Nanoparticles}

\begin{document}

	

\begin{abstract}

The tuning of exchange bias (EB) in nanoparticles has garnered significant attention due to its diverse range of applications. Here, we demonstrate EB in single-phase CoO nanoparticles, where two magnetic phases naturally emerge as the crystallite size decreases from 34.6±0.8 to 10.8±0.9 nm. The Néel temperature ($T_N$) associated with antiferromagnetic ordering decreases monotonically with the reduction in crystallite size, highlighting the significant influence of size effects. The 34.6 nm nanoparticles exhibit magnetization irreversibility between zero-field cooled (ZFC) and field-cooled (FC) states below $T_N$. This irreversibility appears well above $T_N$ with further reduction in size, resulting in the absence of true paramagnetic regime which indicates the occurnace of an additional magnetic phase. The frequency-dependent ac-susceptibility in 10.8 nm nanoparticles suggests slow dynamics of disordered surface spins above $T_N$, coinciding with the establishment of long-range order in the core. The thermoremanent magnetization (TRM) and iso-thermoremanent magnetization (IRM) curves suggest a core-shell structure: the core is antiferromagnetic, and the shell consists of disordered surface spins causing ferromagnetic interaction. Hence, the exchange bias in these CoO nanoparticles results from the exchange coupling between an antiferromagnetic core and a disordered shell that exhibits unconventional surface spin characteristics. 

\end{abstract}

\maketitle

\clearpage

\section{Introduction} 
Exchange bias refers to the shift of the magnetic hysteresis loop of a material along the field axis~\cite{PhysRev.105.904,migliorini,wang2012l,jia2017}. It appears in a system with two different exchange-coupled magnetic phases after cooling in the presence of an external magnetic field. This phenomenon is well-known as conventional exchange bias (CEB) and has been observed in heterostructures consisting of ferromagnetic (FM)-antiferromagnetic (AFM), FM-spin glass, AFM-ferrimagnetic, FM-ferrimagnetic, and FM-FM  phases~\cite{PhysRev.105.904,migliorini,wang2012l,jia2017,tian2016,PhysRevLett.110.107201,domann2016strain,PhysRevLett.79.1130}. Besides heterostructures, CEB has also been reported in nanoparticles due to naturally occurring two magnetic phases, predominantly with an antiferromagnetic ground state where surface spins play a significant role~\cite{PhysRevLett.79.1393,PhysRevLett.101.097206,salabacs2006exchange,PhysRevB.64.174420,huang2018exchange,ahmadvand2010,jagodivc2009}. Notably, a prominent exchange bias effect even in the ZFC condition, known as spontaneous exchange bias (SEB), has been reported in a few systems~\cite{gokemeijer1999,miltenyi1999,saha2007,tomou2006weak}. Until now, SEB has been reported in systems such as IrMn/FeCo bilayers, Ni-Mn-Sn alloys, Si-substituted MnNiSnSi system, BiFeO$_3$-Bi$_2$Fe$_4$O$_9$ nanocomposites~\cite{migliorini,wang2012l,PhysRevLett.110.107201,han2013magnetic,PhysRevB.75.134409,coutrim2018}. The exchange bias effect plays a pivotal role in both fundamental research and numerous applications, including permanent magnets, magnetic resonance imaging, and spintronic devices such as magnetic tunnel junctions, magnetoresistance sensors, and magnetic random-access memory~\cite{peng2020exchange,fang2022electrical,nogues2005exchange}. 

In this context, AFM nanoparticles could be intriguing, given the potential occurrence of two magnetic phases naturally with decrease in their size ~\cite{PhysRevLett.101.097206,salabacs2006exchange,TracyPRB2006}. The magnetic properties of CoO nanoparticles are less explored compared to other oxide nanostructures like NiO since their synthesis is challenging due to the greater thermodynamic stability of the Co$_3$O$_4$ phase, leading to the immediate reducibility of CoO to Co metal~\cite{Vikash_JMMM}. Furthermore, there is no report on SEB and CEB effects in single-phase nanosized CoO, where two exchange-coupled magnetic phases occur naturally with the decrease in crystallite size~\cite{dutta2008room,ghosh2005,santos2021}. This lack of information motivated us to explore the size-dependent magnetic properties of CoO nanoparticles further. Recently, we developed a method for synthesizing separate single-phase face-centered cubic (fcc) and monoclinic phases of reasonably monodispersed CoO nanoparticles with varying crystallite size~\cite{Vikash_JMMM}. The dominant role of trioctylphosphine (TOP) was found to influence the variation in crystallite size during synthesis. The development of new methods for the synthesis of magnetic nanostructures is of particular interest as they have many technological applications~\cite{halder2019,halder2021}. For example, Chen \textit{et al.} demonstrated a significant enhancement in the charge transfer process of BiVO$_4$/CoO$_x$ via Fe doping, which is useful for water splitting~\cite{chen2023boosting}.

In the present study, we present the results of magnetic measurements on rock-salt cubic CoO nanoparticles with varying crystallite sizes ranging from 10.8 nm to 34.6 nm. As the crystallite size decreases, $T\rm_N$ shifts towards a lower temperature. The 10.8 nm nanoparticles exhibit $T\rm_N \sim$ 210 K that is significantly reduced compared to $T\rm_N \sim$ 293 K for bulk CoO. The magnetization and coercivity increase with a decrease in crystallite size. The frequency-dependent ac-susceptibility of 10.8 nm nanoparticles shows prominent dispersion even above $T\rm_N \sim$ 210 K, indicating the glassy response of the surface spins, although there is no shift in $T\rm_N$ with frequency. Isothermal remanent magnetization (IRM) and thermo-remanent magnetization (TRM) measurements reveal the core-shell magnetic behavior of these nanoparticles. IRM and TRM curves are significantly different compared to spin glass, superparamagnet, and diluted antiferromagnet in the field (DAFF)~\cite{PhysRevLett.101.097206,salabacs2006exchange}. They highlight the unconventional nature of surface spins, giving rise to a stronger FM-like response evident from the large value of IRM. These results indicate that spins are aligned antiferromagnetically in the core, whereas the shell consists of disordered surface spins. This core-shell magnetic behavior elucidates the observed EB effects in these nanoparticles.

\section{Experimental details} 

In the present study, we utilized well-characterized single-phase CoO nanoparticles, Co/CoO nanocomposites, and silica-coated CoO nanoparticles, denoted as CoO@SiO$_2$\cite{Vikash_JMMM}. These samples were prepared using thermal decomposition method, and the crystallite size was varied by changing the amount of trioctylphosphine (TOP)\cite{Vikash_JMMM}. Typically, 0.25 ml of preheated TOP at 200 $^{\circ}$C was added to a solution of 2 g of Co(CH$_3$COO)$_2$·4H$_2$O and 8 ml of oleylamine (OA), already degassed at 120–130 $^{\circ}$C for half an hour. The resulting solution was further heated to 230 $^{\circ}$C for 1.5 h under an argon atmosphere. After cooling down to 30 $^{\circ}$C, the reaction product was washed and dried; the sample was denoted as CoO4. Other samples, coded as CoO6 and CoO8, were prepared with TOP concentrations of 3 and 12 ml, respectively, with other conditions remaining the same. Additionally, we prepared an extra sample, CoO10, of Co/CoO nanocomposites with a combination of 8 ml of OA and 12 ml of TOP at 260 $^{\circ}$C in 2 hrs, with other conditions remaining constant. This nanocomposite sample of Co/CoO was prepared to compare the magnetic properties with single-phase CoO nanoparticles since Co/CoO is a prototype example of the exchange bias effect, which is the focus of this paper.

Furthermore, we coated SiO$_2$ on the prepared powder of the smallest-sized nanoparticle sample CoO8 with the specific aim of observing the influence of surface spins on magnetic properties. To functionalize SiO$_2$ on the surface of CoO8 nanoparticles, we employed a standard Stöber method described elsewhere~\cite{verma2017}. In brief, 0.10 g of 10.8 nm nanoparticles were ultrasonicated in a mixture of 40 ml of ethanol, 10 ml of deionized water (DIW), and 1.2 ml of 28 wt. ammonia solution for 1 h. Then, 0.43 mL of tetraethyl orthosilicate (TEOS) was added dropwise to this mixture. After stirring for 6 h, the obtained product was separated by centrifugation for 15 min at 9000 rpm and then washed with DIW. The product was dried at 100 °C and then crushed well to make a fine powder.

\begin{table}[ht]
	\caption{\label{tab:gd2os3si5_cdw_crystalinfo}%
		Synthesis conditions and crystallite size of the CoO nanoparticles.}
	\scriptsize
	\centering
	\begin{tabular}{cccc}
		\hline \\ \\
		Sample & CoO3 & CoO6 & CoO8 \\
		TOP (ml) & 0.25 & 3 & 12  \\
		Reaction temperature (K)&503& 503&503  \\
			Crystallite size, $d$ (nm)& 34.6$\pm$0.8 & 22.9$\pm$0.7& 10.8$\pm$0.9 \\\\
		\hline \\ \\
	\end{tabular}
\end{table}

Laboratory X-ray diffraction (XRD) was performed using a D2 Phaser X-ray diffractometer equipped with a Cu-$K_{\alpha}$ monochromatic source with a wavelength of $\lambda$ = 1.5406~\AA~. Synchrotron radiation XRD measurements were carried out at Beamline station P03, Petra III, Germany, in transmission mode on an image pellet. Transmission electron microscopy (TEM) measurements were performed with a TECHNAI-20-G2 system operated at a 200 kV accelerating voltage. Specimens for these measurements were prepared by drop-casting a well-sonicated solution of a few milligrams of nanoparticles dispersed in approximately 5 ml of ethanol onto carbon-coated TEM grids. X-ray absorption near-edge spectroscopic (XANES) measurements were carried out at BL-1 beamline of the Indus-2 synchrotron radiation source at RRCAT, Indore, India. DC-magnetization measurements were performed using a 7 T SQUID-VSM magnetometer (Quantum Design, USA). The temperature dependence of magnetization $M$(T) curves was recorded in ZFC and FC protocols. In the ZFC protocol, we initially cooled the sample in the absence of an external field down to 5 K. At 5 K, we switched on the magnetic field and recorded the ZFC $M$(T) curve while warming. The sample was then cooled in the presence of the same magnetic field, and the FC $M$(T) curve was recorded during subsequent warming.

The isothermal magnetization, $M-H$ loop for CoO3, CoO8, and CoO10 in ZFC and FC protocols, has been measured. In the ZFC mode, the $M-H$ loop is measured after the sample has been cooled to the measurement temperature in the absence of an applied field. In the FC mode, the $M-H$ loops have been performed while the sample has been cooled at $\mu_0H$ = 7 T. The thermoremanent magnetization (TRM) is recorded after cooling the sample down to 5 K in the presence of different cooling fields. After the temperature stabilizes at 5 K, the applied field is reduced to zero, and the magnetization of the sample has been measured. On the other hand, for the isothermal remanent magnetization (IRM) measurements, the sample is cooled down to 5 K without applying the field. When the temperature reaches and stabilizes at 5 K, the magnetic field is applied momentarily, removed again, and the magnetization of the sample is measured. The frequency-dependent ac susceptibility measurements were carried out using an MPMS XL SQUID magnetometer in the temperature-stable mode.

\section{Results and Discussion} 
\subsection{Structural study}
 
\begin{figure*}[!]
	\includegraphics[scale=0.5]{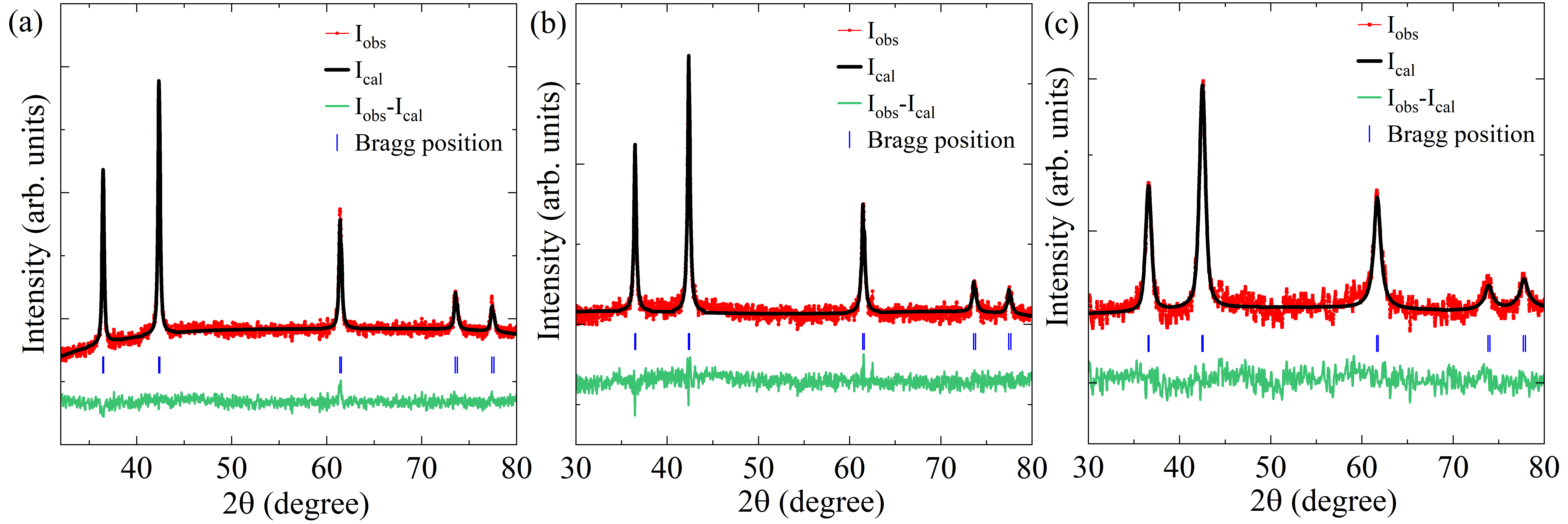}
	\caption{X-ray diffraction patterns along with Rietveld refinement of (a) CoO3, (b) CoO6, and (c) CoO8.}
	\label{fig1}
\end{figure*}

The crystal structure of CoO3, CoO6, and CoO8 was identified using XRD patterns. Figures~\ref{fig1} (a-c) display the laboratory XRD patterns along with the Rietveld refinement of CoO3, CoO6, and CoO8. The five peaks appearing around 36.3$^\circ$, 42.6$^\circ$, 61.4$^\circ$, 73.5$^\circ$, and 77.4$^\circ$ in the XRD pattern of CoO3 correspond to (1 1 1), (2 0 0), (2 2 0), (3 1 1), and (2 2 2) planes, respectively, of the fcc phase of CoO (Fig.~\ref{fig1} (a)). The similar five peaks in the XRD patterns of CoO6 and CoO8 confirm their fcc phase (Figs.~\ref{fig1} (b, c)). Notably, the peaks are broadened from CoO3 to CoO8, indicating a reduction in crystallite size with an increase in the concentration of TOP~\cite{Vikash_JMMM,sharma2020ultralow,sharma2020influence}. The average crystallite size was evaluated using the Debye-Scherrer formula, defined as $L$ = $\frac{K \lambda}{\beta \cos\theta}$, where $L$ is the crystallite size, $\beta$ is the full width at half maximum (FWHM) of the peak, $\theta$ represents the Bragg angle, and $K$ is an anisotropic constant in the range of 0.87–1.0. Here, we used $K$ = 0.9 considering the spherical nanoparticles. The crystallite size is found to be 34.6$\pm$0.8 nm, 22.9$\pm$0.7 nm, and 10.8$\pm$0.9 nm for CoO3, CoO6, and CoO8, respectively, listed in Table~\ref{tab:gd2os3si5_cdw_crystalinfo}. Furthermore, the XRD patterns were analyzed using the Rietveld refinement method implemented with the FullProf Suite package~\cite{Fullprof}. Rietveld analysis reveals that the prepared nanoparticle samples of CoO crystallize in the rock-salt cubic or face-centered cubic (fcc) phase with space group Fm-3m (No. 225). The obtained lattice parameters are \textit{a} = 4.267, 4.265, and 4.258 Å for CoO3, CoO6, and CoO8, respectively. Although these samples, CoO3, CoO6, and CoO8, consist of single-phase CoO nanoparticles, their magnetic properties vary significantly due to distinct crystallite sizes, primarily influenced by the dominant role of finite size and surface effects. As crystallite size decreases, surface effects become more prominent, leading to different magnetic behavior at the surface compared to the core of the nanoparticles, as discussed later.

\begin{figure*}[t]
	
	\includegraphics[width=0.6\columnwidth]{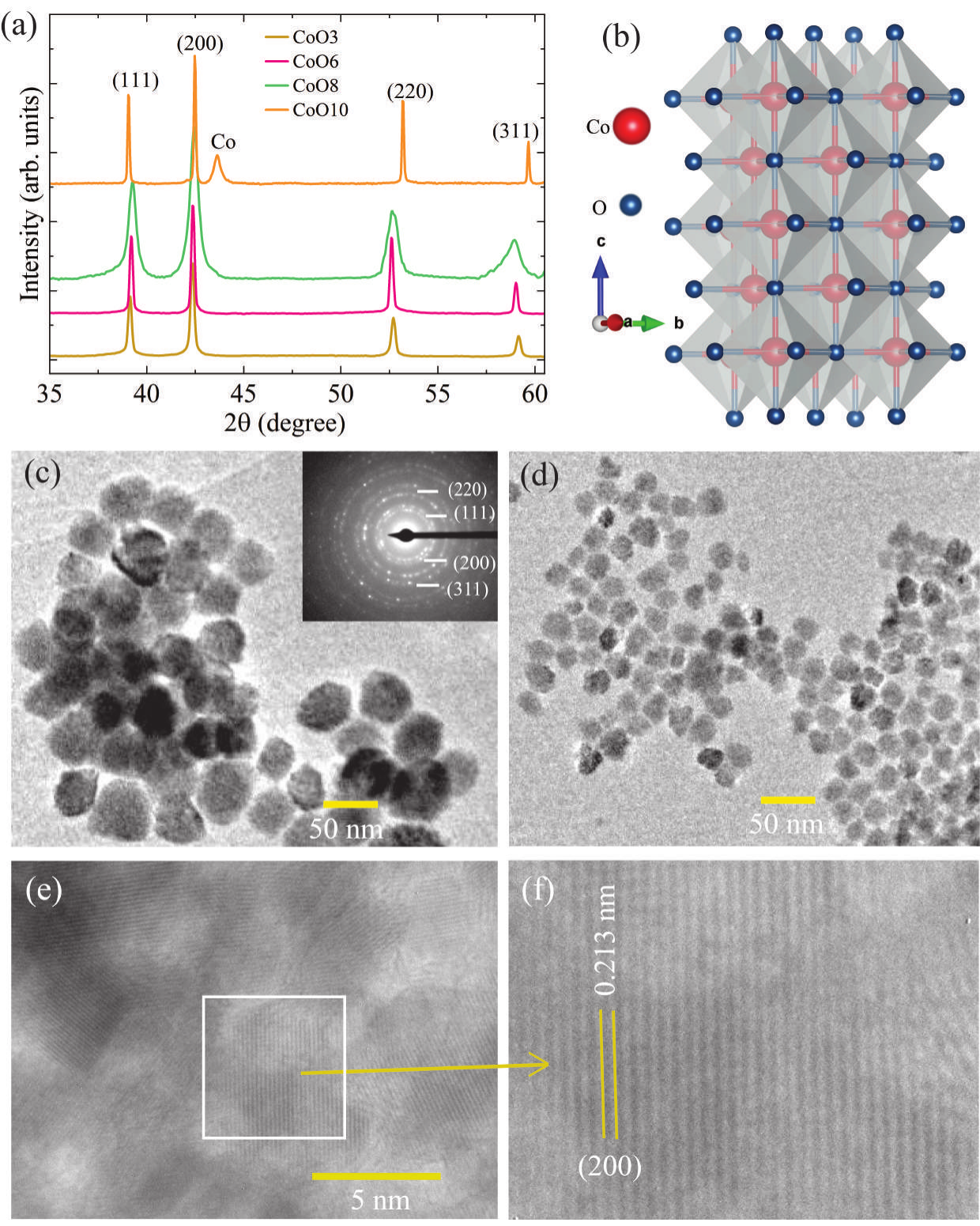}
	\caption{(a) X-ray diffraction patterns of CoO3, CoO6, CoO8 and CoO10 measured using synchrotron radiation, (b) crystal structure of CoO, and transmission electron micrographs of (c) CoO3 and (d) CoO6; inset shows selected area electron diffraction pattern. (e) High-resolution transmission electron micrograph of CoO6 and (f) expanded view of a selected region of (d).}
	\label{fig2}
\end{figure*}

We have also performed synchrotron radiation XRD of CoO3, CoO6, CoO8, and CoO10 (Fig.~\ref{fig2} (a)). The four peaks around 36.4$^{\circ}$, 42.3$^{\circ}$, 61.4$^{\circ}$, and 73.6$^{\circ}$ correspond to (1 1 1), (2 0 0), (2 2 0), and (3 1 1) planes, respectively, for CoO3, CoO6, and CoO8 have been observed. However, the XRD pattern of CoO10 shows an extra peak relative to CoO10, corresponding to Co, indicating that this sample is a composite of Co and CoO. The progressive increase in peak width observed in XRD peaks from CoO3 to CoO8 signifies a corresponding decrease in crystallite size, consistent with findings from the Lab XRD.

The crystal structure of rock-salt cubic CoO is visualized using VESTA~\cite{Vesta} as shown in (Fig.~\ref{fig2} (b)). The crystal structure shows that the effective number of Co atoms is four per unit cell, located at the corners and face centers of the unit cell. The Co$^{2+}$ is bonded to six equivalent O$^{2-}$ atoms to form a mixture of edge and corner-sharing CoO6 octahedra. The Co–O bond lengths are about 2.14 $\rm\AA$. The Co and O atoms have only one site with 4$a$ and 4$b$ Wyckoff positions, respectively.

The TEM micrographs along with selected area electron diffraction (SAED) patterns of CoO3 and CoO6 are depicted in Fig.~\ref{fig2}(c, d). The size distribution of CoO3 and CoO6 is between 10 to 40 nm and 10 to 20 nm with an average particle size around 42 ± 2 and 25 ± 1 nm, respectively. The SAED pattern of CoO3 shows reflections from (1 1 1), (2 0 0), (2 2 0), (3 1 1), and (2 2 2) planes as indicated in the inset of Fig.~\ref{fig2}(c). These reflections correspond to the fcc phase of CoO nanoparticles and are consistent with reflections obtained from XRD data. The high-resolution TEM micrograph of CoO6 shows an interplanar spacing around 0.213 nm corresponding to the (2 0 0) plane (Fig.~\ref{fig2}(e)) which is clearly seen in its expanded view~(Fig.~\ref{fig2}(f)).

\subsection{X-ray absorption near edge spectroscopy}

\begin{figure*}[t]
	\includegraphics[scale=1]{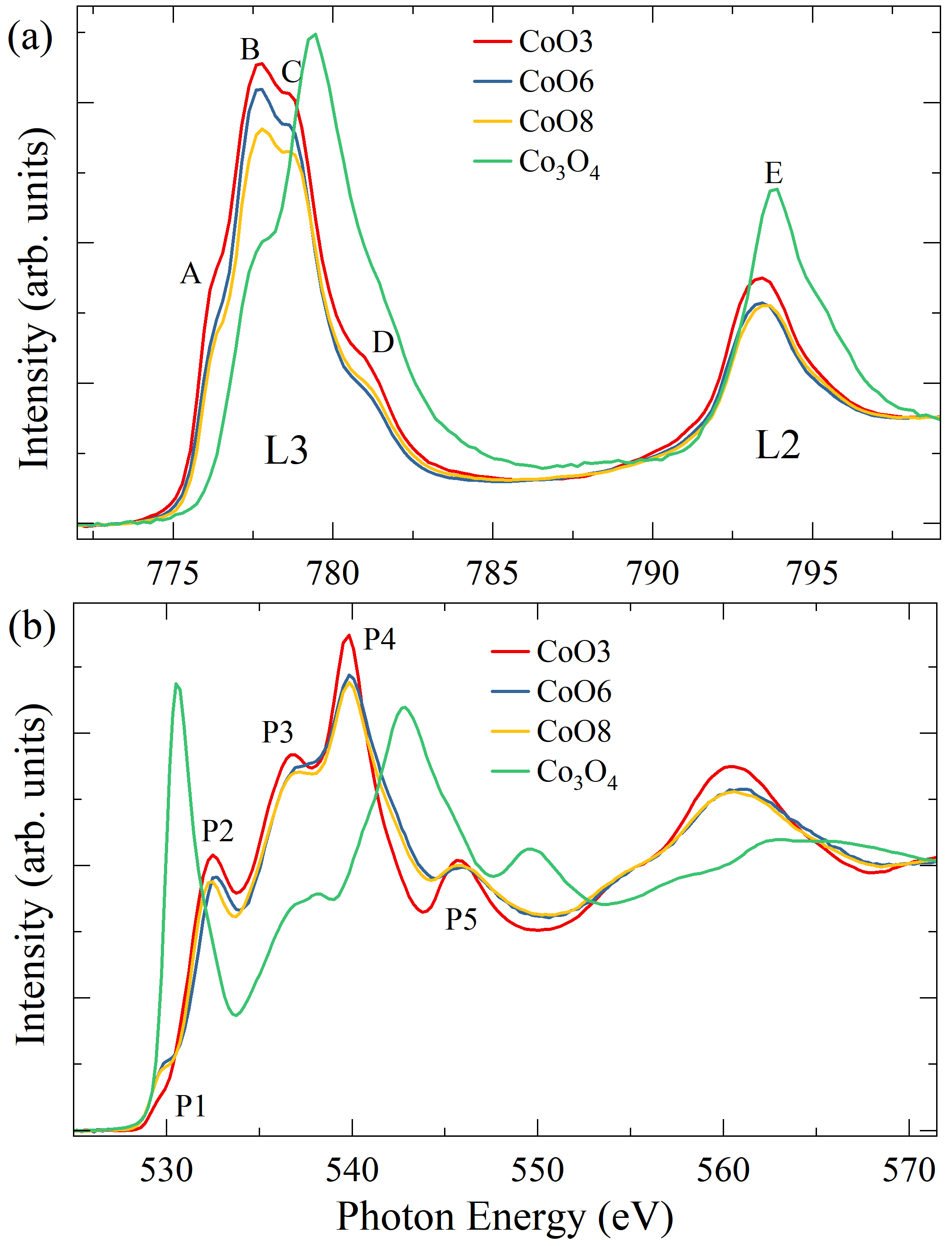}
	\caption{X-ray absorption near edge spectra of (a) Co 2p and (b) O 1s for CoO3, CoO6 and CoO8 along with bulk Co$_3$O$_4$ as reference.}
	\label{fig3}
\end{figure*}

We employed XANES to determine the valence state of the Co ion in the prepared samples. We present the normalized XANE spectra of Co 2p and O 1s for CoO3, CoO6, and CoO8, along with bulk Co$_3$O$_4$ as shown in Fig.~\ref{fig3} (a). The Co $L$-edge splits into two main features due to core level spin-orbit coupling, denoted as the lower energy $L3$ and the higher energy $L2$, which are assigned to the transitions from Co 2p$\frac{3}{2}$ ($L3$) and 2p$\frac{1}{2}$ ($L2$) electrons into the empty 3d states, respectively.The energy separation between $L3$ and $L2$ edges in CoO3 is found to be about 15.7 eV, well-matched with the value obtained from photoelectron spectroscopy (XPS)\cite{Vikash_JMMM}. Furthermore, the Co $L3$ edge is split into four features at around 776.1 (A), 777.6 (B), 778.8 (C), 781.0 (D), mainly due to electron-electron interactions\cite{hibberd2015}. Therefore, the Co L edge of CoO3 exhibits five main features near 776.1 (A), 777.6 (B), 778.8 (C), 781.0 (D), and 793.4 (E) eV, between 772.5 and 798.5 eV, which are in agreement with earlier reports~\cite{hibberd2015,magnuson2002}. CoO6 and CoO8 also show similar five features. The XANE spectrum of Co$_3$O$_4$ shows mainly three peaks around 777.7, 779.4, and 794.0 eV, in line with an earlier report~\cite{hibberd2015}. The Co $L3$ edge of Co$_3$O$_4$ has only two main features. Note that the multiplet structure of the $L3$-edge is highly sensitive to the coordination of the Co ion. The Co 2p spectrum of CoO is due to octahedral Co$^{2+}$ crystal field splitting and is significantly different from Co$_3$O$_4$, in which both octahedral Co$^{2+}$ and tetrahedral Co$^{3+}$ coordinations are present.

The 2p orbitals of the oxygen ligand are involved in the bonding configuration with Co metal ions. Therefore, dipole-allowed 1s to 2p transitions of O 1s can directly provide information about the oxygen charge state and reflect the partial density of O 2p unoccupied electronic states in the conduction band~\cite{galakhov2006,soriano1999}. We measured the O 1s XANE spectra of CoO3, CoO6, and CoO8 along with bulk Co$_3$O$_4$, as shown in Fig.~\ref{fig3}(b).The O 1s spectra of these samples exhibit mainly five features at around 529.5 (P1), 532.3 (P2), 536.5 (P3), 539.8 (P4), and 549.5 (P5) eV, ranging between 525.5 to 552.5 eV. The peaks P2 to P5 in these nanoparticles are in close agreement with earlier reports~\cite{soriano1999,hibberd2015,magnuson2002}. However, an additional pre-edge feature, P1, is found compared to bulk CoO, and it becomes more pronounced with the decrease in crystallite size. This feature indicates that the electronic structure is modified in nanoparticles compared to the bulk counterpart.
The features between 530 and 535 eV correspond to transitions to unoccupied electronic states of O 2p character hybridized with Co 3d states~\cite{soriano1999}. Above 535 eV, features in XANE are mainly associated with transitions to O 2p character hybridized with Co 4s, 4p states~\cite{soriano1999}. The O 1s character of Co$_3$O$_4$ is significantly different from that of CoO nanoparticles. For example, a sharp peak that appears in Co$_3$O$_4$ is absent in CoO nanoparticles. These results indicate that the bonding configuration of O ions with Co metal ions, and also valence states, are different in CoO nanoparticles and Co$_3$O$_4$. Therefore, the XANES results of CoO3, CoO6, and CoO8 clearly show that these nanoparticles exhibit a pure Co$^{2+}$ valence state, which is in line with XRD data

\subsection{Magnetic Properties}
\begin{figure*}[t]
	\centering
	\includegraphics[scale=0.8]{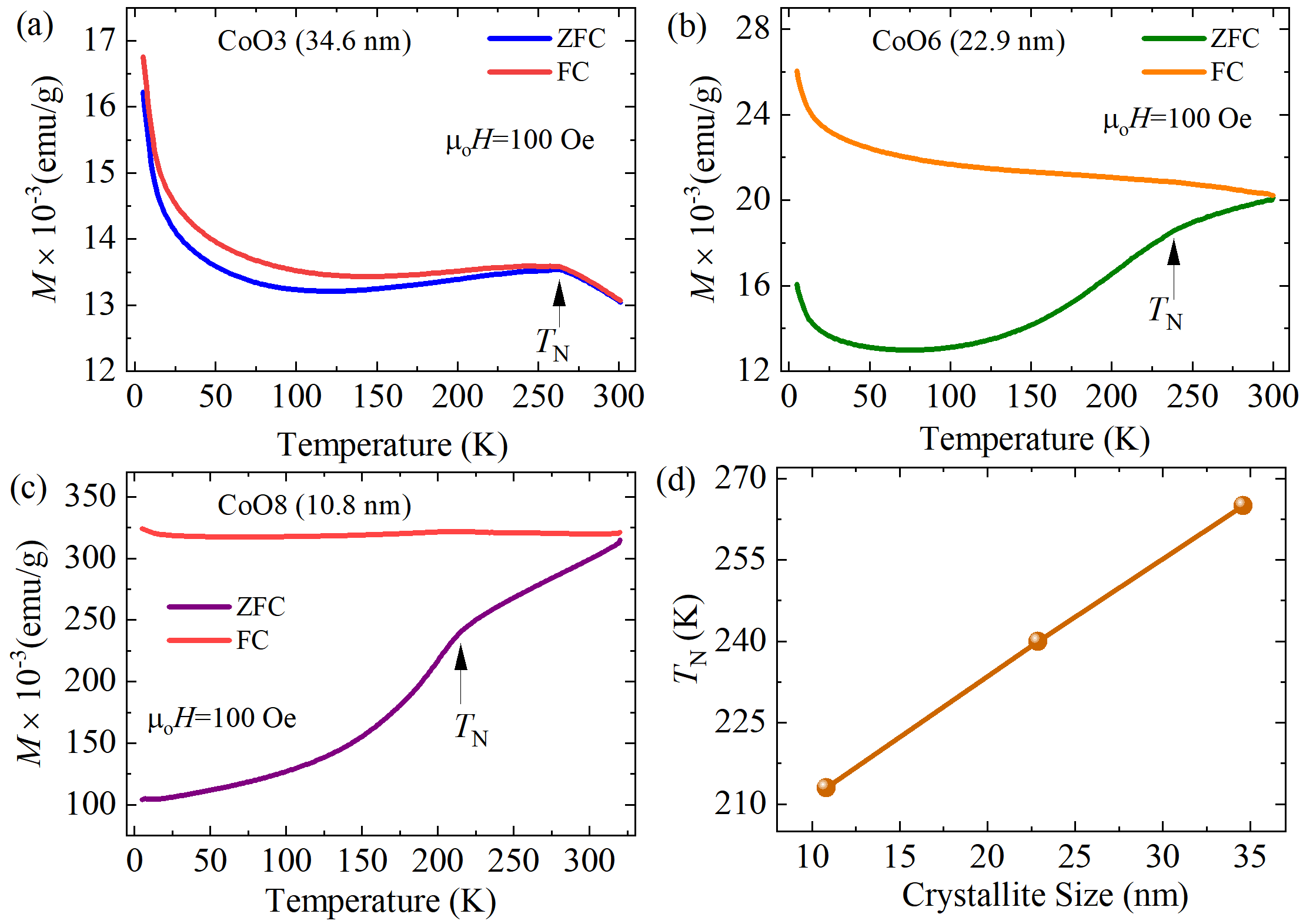}
	\caption{Temperature dependent magnetization (M) of (a) CoO3, (b) CoO6, and (c) CoO8 measured in zero field-cooled (ZFC) and field-cooled (FC) protocols at an applied magnetic field of $\mu_oH$ = 100 Oe; the Néel temperature ($T\rm_N$) is indicated by the arrow. (d) Variation in $T\rm_N$ as a function of crystallite size.} 
	\label{fig4}
\end{figure*}

Figure~\ref{fig4}(a-c) shows the temperature-dependent magnetization of CoO3, CoO6, and CoO8, respectively, measured in the ZFC and FC protocols under an applied dc field of $\mu_0H$ = 100 Oe. In the case of CoO3, the ZFC and FC magnetization curves gradually increase as the temperature is reduced from 300 K. Both curves exhibit a shallow hump at $T_N \sim$ 265 K, indicating the formation of an antiferromagnetic state. It is noteworthy that $T_N \sim$ 265 K in CoO3 (34.6 nm) nanoparticles is significantly reduced compared to $T_N \sim$ 293 K of bulk CoO~\cite{TracyPRB2006}. This suggests that the finite size effect starts to play a crucial role in CoO nanoparticles at this size~\cite{ambrose1996finite}. The temperature where the first derivative of the ZFC magnetization versus temperature data exhibits a maximum was taken as $T_N$. The FC curve follows the ZFC curve above $T_N$, but they start to bifurcate below $T_N$. The thermomagnetic irreversibility is defined as $\Delta M = (M_{FC} - M_{ZFC})$, where $M_{FC}$ and $M_{ZFC}$ are measured magnetization in FC and ZFC, respectively, and it increases with a decrease in temperature (Fig.\ref{fig4}(a)). Further decrease in temperature below $T_N$, the ZFC magnetization curve decreases until the minimum temperature ($T_{min} \sim$ 125 K) and then gradually increases. The $M_{FC}$ curve also shows a qualitatively similar temperature dependence. The sharp rise of magnetization at low temperatures is not a typical characteristic feature of a long-range antiferromagnetic state, and in this case, it may be due to moments on the defect sites in CoO nanoparticles\cite{TracyPRB2006}.

The $M\rm_{FC}$ and $M\rm_{ZFC}$ curves of CoO6, with an average crystallite size of approximately $d = 22.9$ nm, exhibit a more interesting magnetic response (Fig.\ref{fig4}(b)). The hump associated with the AFM transition has shifted down to $T\rm_N \sim 240$ K and has become broader compared to CoO3. A finite bifurcation between $M\rm_{FC}$ and $M\rm_{ZFC}$ curves now exists above $T\rm_N$ and persists even at 300 K. Below $T\rm_N \sim 240$ K, the ZFC curve decreases rapidly, exhibits a minimum around $T\rm_{min} = 75$ K, and then increases again as the temperature further reduces. Fig.\ref{fig4}(c) displays the $M\rm_{FC}$ and $M\rm_{ZFC}$ curves of CoO8 with a crystallite size of 10.8 nm. In CoO8, the ZFC curve exhibits a broad hump associated with antiferromagnetic ordering at $T\rm_N \sim$ 213 K. However, $T\rm_N$ is weak and therefore not clearly visible in the FC magnetization curve of CoO8. Notably, CoO8 does not exhibit a sharp increase in magnetization at low temperatures, as observed in the case of CoO3 and CoO6. The bifurcation between ZFC and FC magnetization persists even at 320 K. It is worth noting that thermomagnetic irreversibility between ZFC and FC magnetizations in CoO6 and CoO8 starts well above $T\rm_N$, and the bifurcation temperature increases with a decrease in crystallite size (Fig.\ref{fig4}(b,c)). This indicates that smaller-size nanoparticle samples are not in a paramagnetic state at room temperature, possibly due to disordered surface spins. However, a bigger size sample, CoO3, shows the paramagnetic regime in which both ZFC and FC curves merged above $T\rm_N$ (Fig.\ref{fig4}(a)). The $T\rm_N$ decreases in a linear fashion with the decrease in crystallite size as depicted in Fig.\ref{fig4}(d), which is consistent with earlier reports\cite{ghosh2005,Zheng2005}. This indicates the dominant role of size effect on the antiferromagnetic transition.

\begin{figure*}[t]
	\centering
	\includegraphics[scale=0.65]{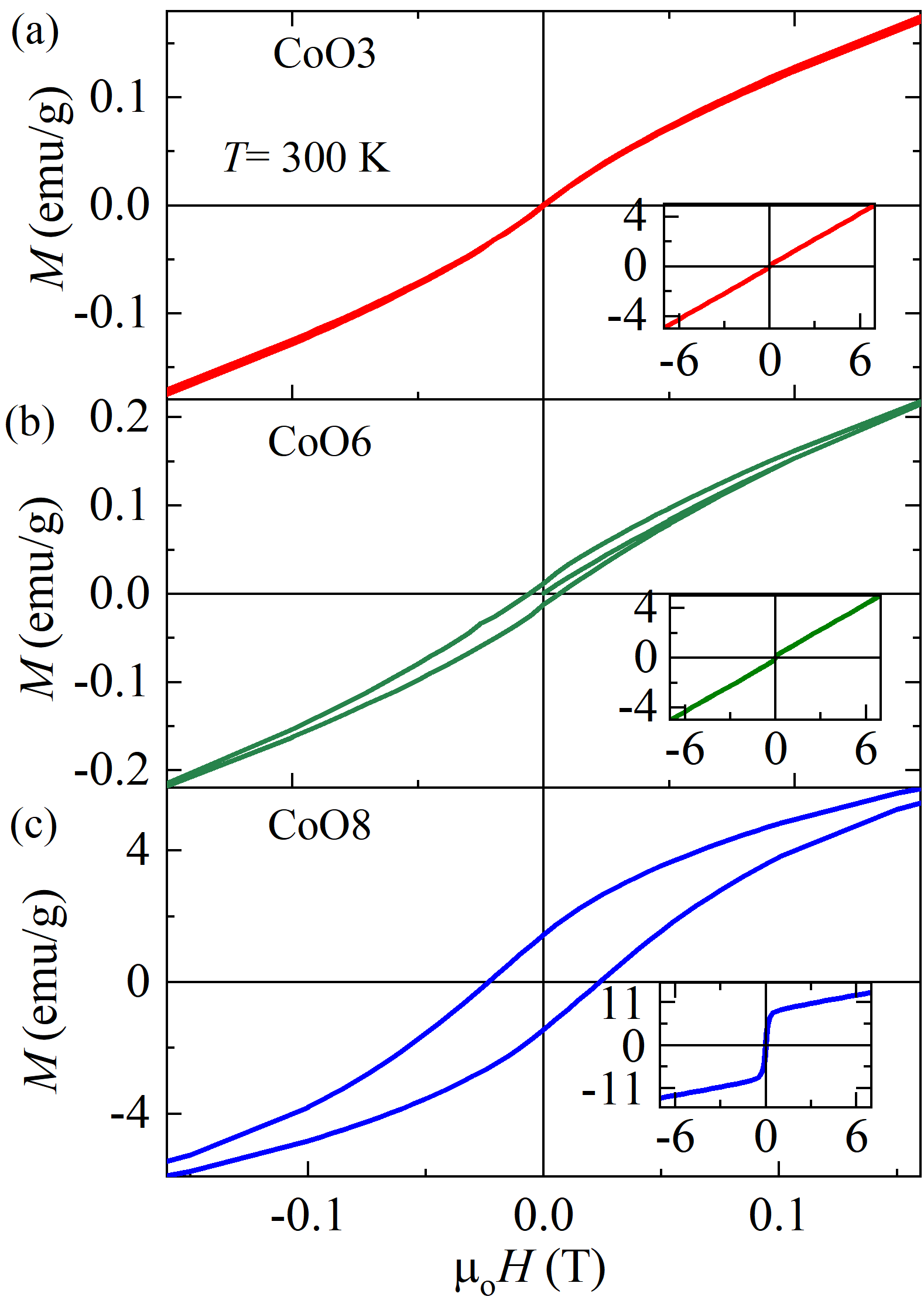}
	\caption{Magnetization (M) versus applied magnetic field ($\mu_oH$) curves  of (a) CoO3, (b) CoO6 and (c) CoO8 in the selected range of $\mu_oH$ at 300 K; insets (a-c) show their $M-H$ curves between -7 T to 7 T.} 
	\label{fig5}
\end{figure*}

To further characterize the magnetic state at 300 K, we have recorded the magnetic field dependence of magnetization ($M-H$ loops) of CoO3, CoO6, and CoO8 at 300 K (Fig.\ref{fig5}). In the case of CoO3, the $M-H$ loop does not show any hysteresis, suggesting the absence of any ferromagnetic-like correlation at 300 K and at this crystallite size. However, CoO3 is still not ideally paramagnetic at 300 K, as evident from the nonlinear variation of M with H (Fig.\ref{fig5}(a)). Furthermore, with a decrease in the crystallite size to 22.9 nm in CoO6, clear hysteresis has been observed near the origin at 300 K (Fig.\ref{fig5}(b)). In furtherance, a prominent hysteresis has been observed at 300 K for CoO8 (Fig.\ref{fig5}(c)). The coercive field ($H_C$) for CoO6 and CoO8 is found to be about 20 and 610 Oe, respectively. The prominent hysteresis indicates a ferromagnetic-like magnetic component in CoO6 and CoO8 nanoparticles even at 300 K, which is much above their respective Néel temperatures. Notably, the $M-H$ curves of the nanoparticles at 300 K are symmetric with respect to the origin, i.e. there is no exchange bias effect. The exchange bias appears in exchange-coupled systems with two different magnetic phases due to additional unidirectional anisotropy. At 300 K, only a ferromagnetic-like component is present, but antiferromagnetic order emerges below 240 K and 213 K for CoO6 and CoO8, respectively.

As particle size decreases, both finite size and surface effects play crucial roles in modifying the magnetic behavior of a material. The increase in dangling bonds and local symmetry breaking leads to spin disorder and frustrations due to the finite size effectcite{Winkler2005}. Surface effects result in a reduced coordination of exchange-coupled spins, giving rise to site-specific surface anisotropy~\cite{Winkler2005}. Consequently, profoundly different magnetic behaviors are observed between the core and the shell in nanomaterials, where the core behaves similarly to the bulk material, while the shell or surface layer may exhibit characteristics such as spin-glass, superparamagnetism, or diluted antiferromagnetism (DAFF)~\cite{PhysRevLett.101.097206,salabacs2006exchange}. This behavior of surface spins is mainly due to uncompensated surface spins~\cite{dutta2008room,rinaldi2014}. These surface spins have been reported earlier to produce a ferromagnetic response superimposed upon the antiferromagnetic response arising from the core of the nanoparticles~\cite{Winkler2005,dutta2008room,rinaldi2014,BenitezPRB}.

\begin{figure*}[t]
	\includegraphics[scale=0.65]{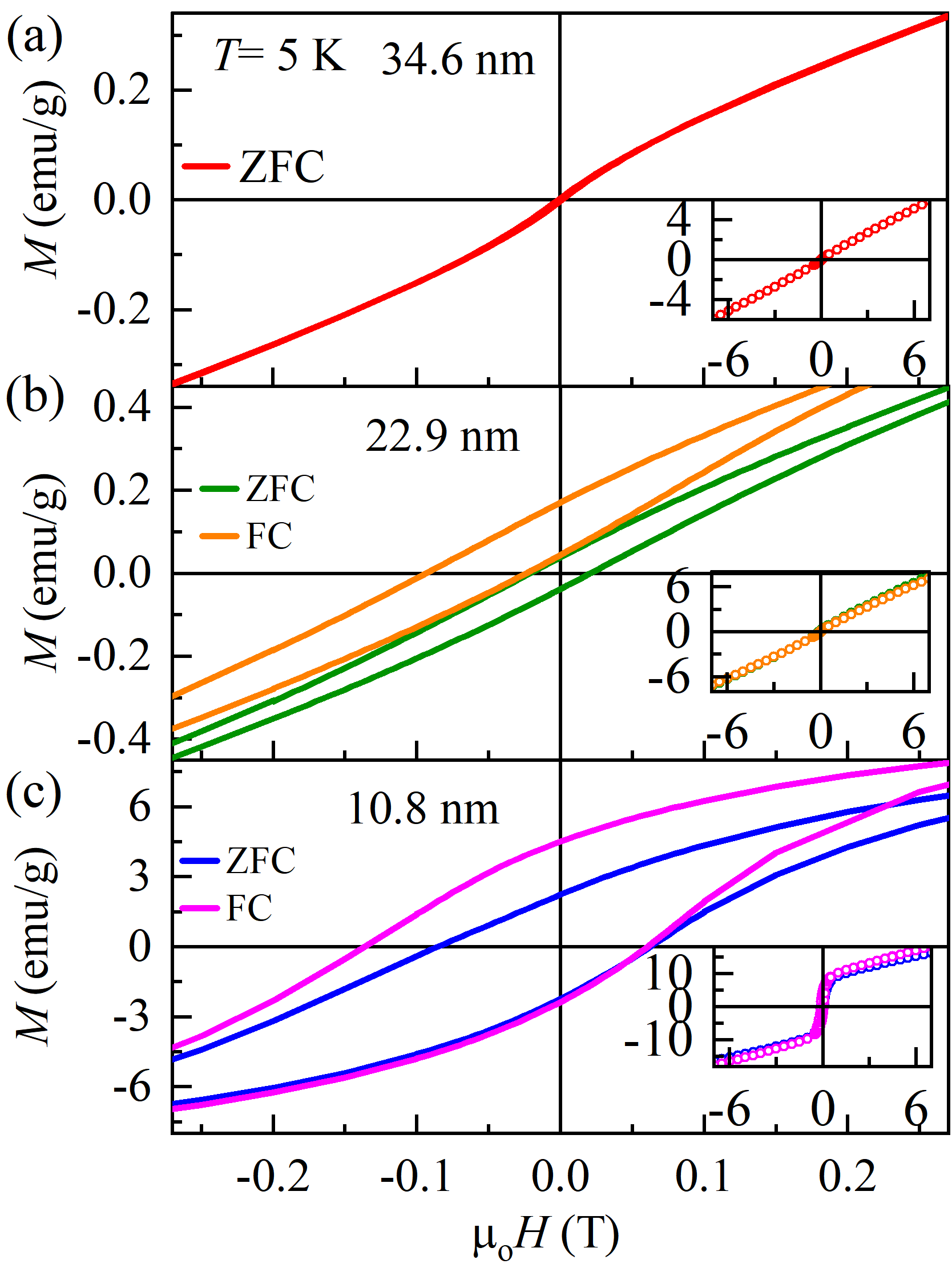}
	\caption{Magnetization (M) versus applied magnetic field ($\mu_oH$) curves of (a) CoO3 measured in zero field-cooled (ZFC) protocol, and (b) CoO6 and (c) CoO8 measured in ZFC and FC protocols in the selected range of $\mu_oH$ at 5 K; insets (a-c) show their $M-H$ curves between -7 T to 7 T.}
	\label{fig6}
\end{figure*}

The M versus H curves of CoO3, CoO6, and CoO8 at 5 K, measured after the samples have been cooled from 400 K, assuming a paramagnetic regime in the absence of an external field, are shown in Fig.\ref{fig6}. In the case of CoO3, the ZFC $M-H$ curve at 5 K does not show any hysteresis at the origin, which is an indication of the prevalence of AFM interactions (Fig.\ref{fig6}(a)). As the particle size reduces, a prominent hysteresis loop is observed, and the coercive field monotonically increases with the reduction in the particle size. The $M-H$ curves of CoO3 and CoO8 are nonlinear and resemble an $S$ shape, without any sign of saturation up to $\mu_0H$ = 7 Tesla (Fig.~\ref{fig6}(b,c)). The $M-H$ loops of CoO6 and CoO8 exhibit coercive fields of $H_C$ = 340 and 1770 Oe, respectively, which are significantly larger than the $H_C$’s = 20 and 610 Oe, respectively, found at 300 K.

Most importantly, the hysteresis loop of the smallest average size CoO8 is asymmetric around the origin. It shifts horizontally while the sample is cooled in the absence of any external field; i.e., CoO8 exhibits SEB at 5 K. The EB field is calculated as $H_{EB} = |(H_{C1}+H_{C2})|/2$, where $H_{C1}$ and $H_{C2}$ are the coercive fields of the shifted hysteresis loop. In this case, a ZFC EB field of $H_{SEB} \approx 120$ Oe is observed. Usually, exchange bias arises in systems comprising antiferromagnetic and ferromagnetic phases, and at the interface of two magnetic phases, wherein a small fraction of spins at the interface of two magnetic phases gets pinned while the sample is cooled in the presence of a finite magnetic field, causing a unidirectional anisotropy, and the hysteresis loop of the whole system shifts with respect to the origin. However, it is now known that besides AFM/FM heterostructures, other composite structures, such as AFM/spin glass, AFM/ferrimagnet, FM/ferrimagnet, etc., can also give rise to such EB~\cite{PhysRev.105.904,migliorini,wang2012l,jia2017,tian2016,PhysRevLett.110.107201,domann2016strain,PhysRevLett.79.1130}. In the case of nanoparticles with an antiferromagnetic ground state, surface spins may give rise to complicated magnetic responses, including weak ferromagnetic, spin glass, two-dimensional dilute antiferromagnetic in a field~\cite{PhysRevLett.101.097206}. In the present case, a small fraction of the spins at the interface of the antiferromagnetic core and disordered surface spins (shell) gets pinned, giving rise to the SEB.

We have also recorded the $M-H$ loop at 5 K in the FC condition to probe the conventional exchange bias (Fig.\ref{fig6}(b, c)). The sample has been cooled to 5 K from 400 K in the presence of $\mu_0H$ = 7 T, and the field is cycled between $\pm$ 7 T, which is sufficient to obtain technical saturation and hence rules out the possibility of the minor hysteresis loop. In this case, we have recorded the $M-H$ loops of only CoO6 and CoO8, which show prominent hysteresis at 5 K in ZFC (Fig.\ref{fig5}(b, c)). Under the FC condition, an enhancement in coercivity as compared to the ZFC protocol has been found in the CoO nanoparticles (Fig.\ref{fig6}(b, c)). The hysteresis loop of CoO8 at 5 K exhibits a larger EB of around $H_{CEB} \approx 381$ Oe in FC at 5 K. In contrast, although CoO6 does not show any exchange bias in the ZFC condition, it shows $H_{CEB} \approx 350$ Oe in FC at 5 K. In this context, we note the reports of Lierop \textit{et al.} of SEB and CEB of around 75 and 150 Oe, respectively, in Ni${80}$Fe${20}$/Co$_3$O$_4$ thin film\cite{PhysRevB.75.134409}, and Takano \textit{et al.} estimated and observed CEB of $H_{CEB} \sim 107$ Oe on Ni${81}$Fe${19}$/CoO bilayers~\cite{PhysRevLett.79.1130}. The value of CEB in the CoO nanoparticles is larger compared to earlier reports on Ni${81}$Fe${19}$/CoO bilayers and Ni${80}$Fe${20}$/Co$_3$O$_4$ thin film~\cite{PhysRevB.75.134409,PhysRevLett.79.1130}.

\begin{figure*}[t]
	\centering
	\includegraphics[scale=0.65]{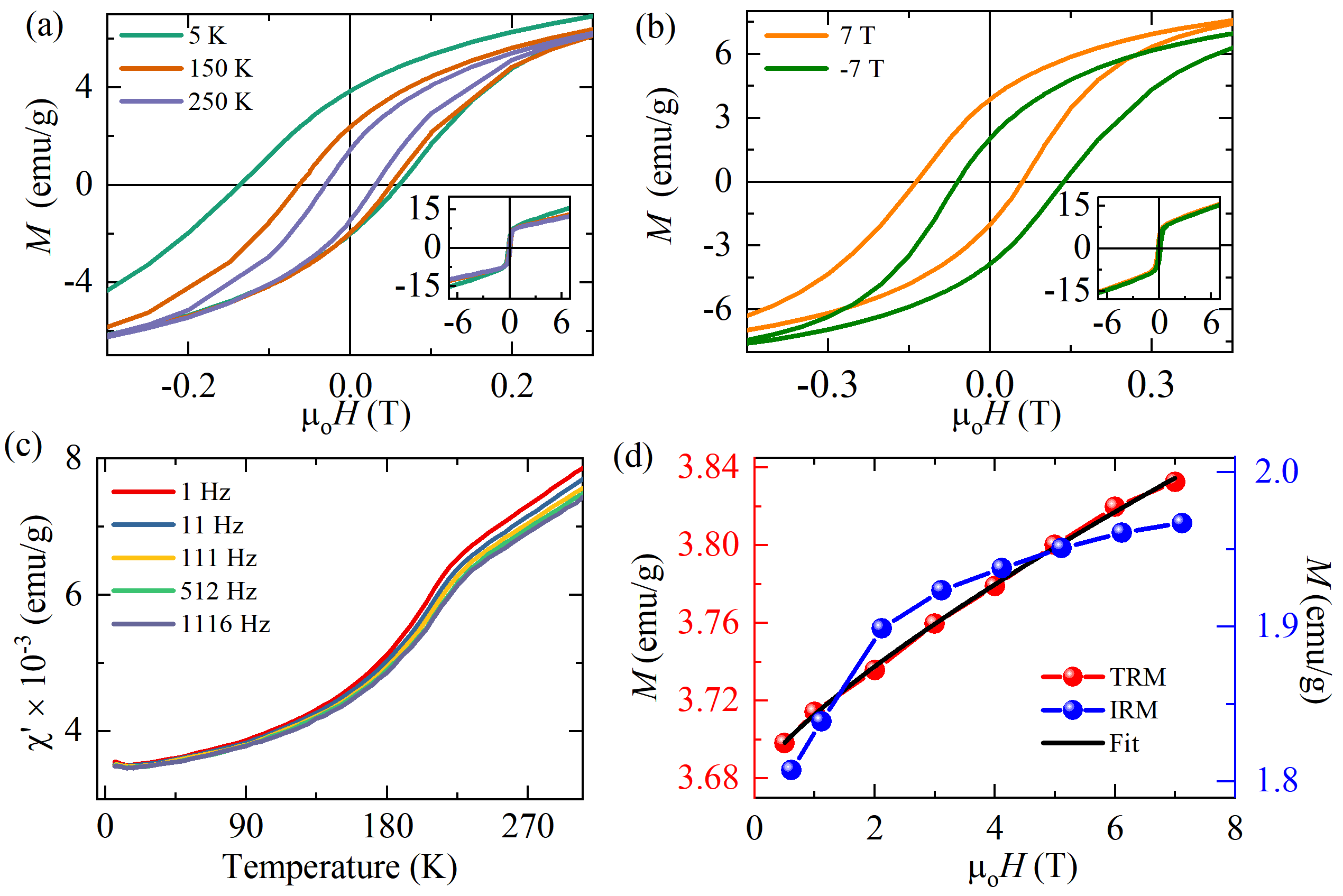}
	\caption{Magnetization (M) versus applied magnetic field ($\mu_oH$) curves of (a) CoO8 under field cooled (FC) protocol at 5 K, 150 K, and 250 K and (b) $M-H$ of CoO8 measured under field cooled (FC) in -7 T and +7 T in selected field range of $\mu_oH$; insets (a, b) show the $M-H$ loops between -7 T to 7 T. (c) Real-part of AC susceptibility ($\chi$) of CoO8 at different frequencies, and  (d) thermoremanent magnetization (TRM); black line shows the power law fit of TRM, and isothermoremanent magnetization (IRM) versus applied field ($\mu_oH$) curves of CoO8.}  
	\label{fig7}
\end{figure*}

Now, to see the role of the antiferromagnetic state behind the EB, we have measured the field-cooled $M-H$ loops of CoO8 at 150 and 250 K, which respectively lie below and above the $T_N$ = 210 K. Fig.\ref{fig7}(a) shows $M-H$ loops at 5 K, 150 K, and 250 K. Interestingly, the magnitude of the shift in the hysteresis i.e., $H_{CEB}$ at 150 K is reduced compared to 5 K. On the other hand, the hysteresis loop at 250 K is symmetric, i.e., EB disappears above $T_N$. However, we would like to reiterate that the bifurcation between the ZFC and FC curves in CoO8 exists even at 300 K. This suggests that the magnetic state above $T_N \sim 210$ K is not truly paramagnetic, and the surface spins give rise to a ferromagnetic-like response. Fig.\ref{fig7}(b) shows the M-H loops when the sample is cooled in +7 T and -7 T. The shift in the hysteresis is found to be nearly equal and opposite direction, which confirms that the asymmetry is an intrinsic characteristic of the sample and not related to the instrument. These results indicate that the EB in the nanoparticles appears due to the presence of an antiferromagnetic state in the core and the disordered shell that gives rise to a net moment in the field.

In this context, it remains essential to investigate the characteristics of the surface spins, which evidently play a crucial role in the magnetic response of the nanoparticles. Therefore, we recorded the frequency-dependent ac susceptibility in the frequency range of 1 to 1116 Hz in the presence of a small ac field of $\mu_0H_{ac} = 3$ Oe. The data is shown in Fig.\ref{fig7}(c). The real part of the ac susceptibility data shows a broad hump around 210 K, associated with the long-range antiferromagnetic arrangement of the spins at the core. This feature aligns well with the temperature-dependent dc magnetization measured in ZFC (Fig.\ref{fig4}(c)). The interesting point here is that the ac susceptibility exhibits significant frequency dependence, which is not expected, in general, in systems with long-range magnetic order. Also, note that the dispersion exists even above $T_N$ and persists up to the highest measured temperature of 320 K. This indicates that the surface spins have a distribution of the timescale and give rise to the metastable magnetic response.

To further characterize the magnetic response of the surface spins, we recorded IRM and TRM measurements, which are useful to gain some understanding of the nature of the surface spins~\cite{PhysRevLett.101.097206,salabacs2006exchange}. The data is shown in Fig.\ref{fig7}(d). The TRM curve (red filled circle) shows a monotonic increase with no sign of saturation up to $\mu_0H= 7$ T. It doesn't show any maximum at low field, which is believed to be characteristic of spin-glass behavior\cite{tholence1974,BinderRMP1986}. Moreover, significant coercivity or loop opening even at 300 K is not expected in a superparamagnetic system. The IRM curve (blue-filled circle), in contrast, shows a completely different behavior compared to TRM. It monotonically increases with a hump-like feature around $\mu_0H=3$ T and nearly saturates at $\mu_0H=7$ T. Thus, the shape of the IRM and TRM curves, and such a higher value of IRM, cannot support a spin-glass, SPM, or DAFF behavior of these nanoparticles~\cite{PhysRevLett.101.097206}, and hence their possibility can be excluded. We fitted the TRM curve using a power law, which varies as TRM $\propto H^\nu$, where $\nu = 0.77$. Bulk-DAFF is characterized by a virtually zero IRM and a monotonically increasing TRM with an exponent $\nu = 3.056$\cite{PhysRevLett.101.097206}. These CoO nanoparticles, i.e., CoO8 sample, show the exponent $\nu = 0.77$, which neither matches with bulk nor 2D finite-size DAFF. Note that the unambiguously higher value of IRM compared to earlier studies\cite{PhysRevLett.101.097206,salabacs2006exchange} on nanostructures also discards the presence of the DAFF state. From the present set of data, we argue that the shell of nanoparticles shows unconventional nature, possibly ferromagnetic-like interactions rather than spin-glass, superparamagnet, or DAFF behavior as reported earlier~\cite{PhysRevLett.101.097206,salabacs2006exchange}. Therefore, IRM/TRM data confirm the core-shell magnetic behavior of the CoO nanoparticles. Their core behaves as regular AFM, as evident from the appearance of $T_N$, and the shell exhibits unconventional glassy features that give rise to FM interaction. We believe that TRM contribution is possibly responsible for CEB in these nanoparticles, and the unambiguous higher value of IRM suggests that SEB originates due to naturally occurring two magnetic phases after zero-field cooling of the sample.

In order to understand the effect of surface spin disorder on EB, we have prepared and studied the magnetic behavior of two composite systems:

(i) Co/CoO nanocomposite, coded as CoO10, which nearly consists of 20~\% of Co and 80~\% of CoO nanoparticles as per the Rietveld analysis of X-ray diffraction.

(ii) CoO8, i.e., 10.8 nm CoO nanoparticles coated with diamagnetic silica (SiO$_2$).

\begin{figure*}[t]
	\centering
	\includegraphics[scale=0.65]{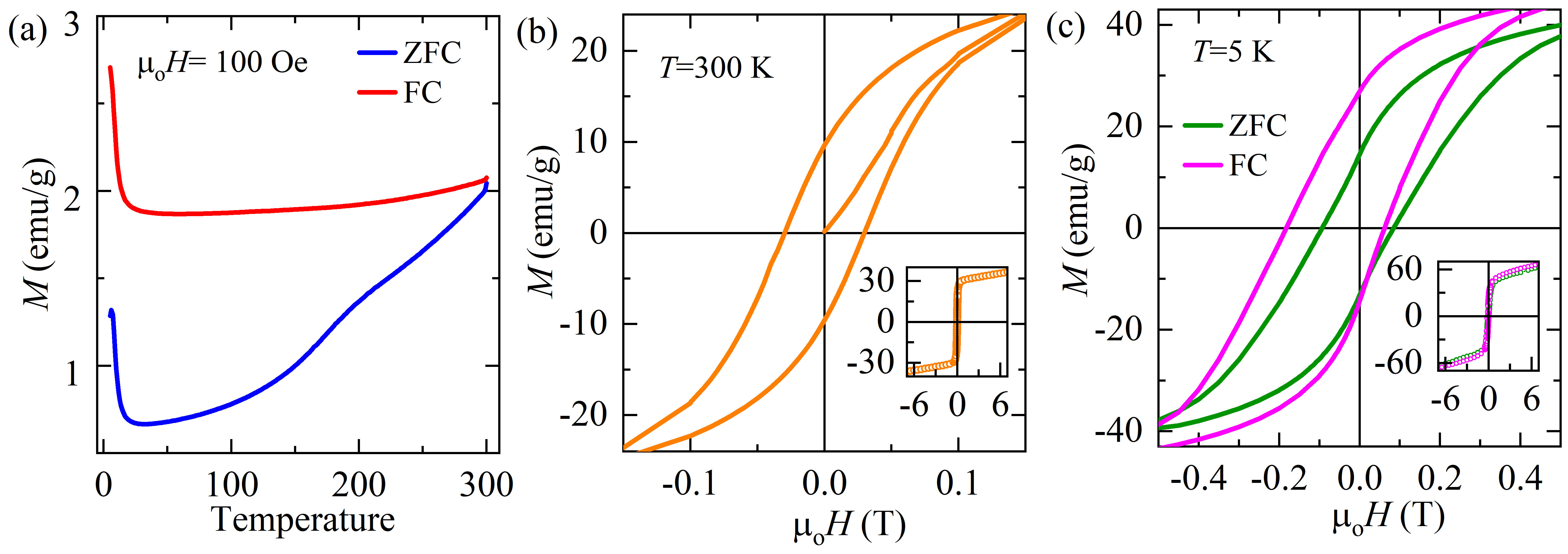}
	\caption{Temperature-dependent magnetization (M) of (a) CoO10 measured in zero field-cooled (ZFC) and field-cooled (FC) protocols at an applied magnetic field ($\mu_oH$) of 100 Oe. Magnetization versus field curves of CoO10 at (b) 300 K and (c) at 5 K measured in ZFC and FC protocols in the selected range of $\mu_oH$; insets (b,c) show the $M-H$ loops between -7 T to +7 T.} 
	\label{fig8}
\end{figure*}

Figure~\ref{fig8}(a) shows the temperature-dependent magnetization of CoO10, which is the composite system of Co and CoO where the average crystallite size of CoO nanoparticles is about 11 nm. The distinct $M-T$ curve of CoO10 compared to CoO8 can be seen, and the value of the magnetization of CoO10 is very high as compared to CoO8.

However, crystallite size is nearly equal in both samples, and measurements were performed in the same applied field of 100 Oe. This extra magnetic moment in CoO10 is mainly attributed to the presence of Co, which is supposed to be ferromagnetic in this temperature range. The bifurcation between the ZFC and FC curves persists up to room temperature. A shallow hump can be observed associated with the Néel temperature of the CoO nanoparticles around $T_N \sim$ 176 K. At low temperatures, a sharp rise in magnetization is observed. The $M-H$ loop at 300 K for CoO10 is shown in Fig.\ref{fig8}(b). It shows a prominent hysteresis loop with a coercivity of $H_c \approx$ 590 Oe. At T = 5 K, the $M-H$ loop is symmetric with respect to the field axis, which means the ZFC $M-H$ loop does not exhibit SEB like in CoO8 (Fig.\ref{fig8}(b)). The $H_C$ of 1750 Oe found at 5 K when measured under ZFC protocol is significantly increased compared to 300 K. To see whether it shows conventional exchange bias, we measured the field-cooled $M-H$ of CoO10 at 5 K in FC at 7 T that shows a clear shift in hysteresis with $H_{EB}$ of 425 Oe (Fig.\ref{fig8}(c)). Thus, Co/CoO nanocomposites show CEB but not SEB, which is in line with an earlier report\cite{skumryev2003}. The CEB in CoO10 is much larger compared to CoO8. The above observation appears to indicate that surface spin plays a crucial role behind the appearance of SEB in nanoparticles, which may not be observed in the conventional system like Co/CoO in which a significant ferromagnetic contribution leads to CEB.

\begin{figure*}[t]
	\centering
	\includegraphics[scale=0.8]{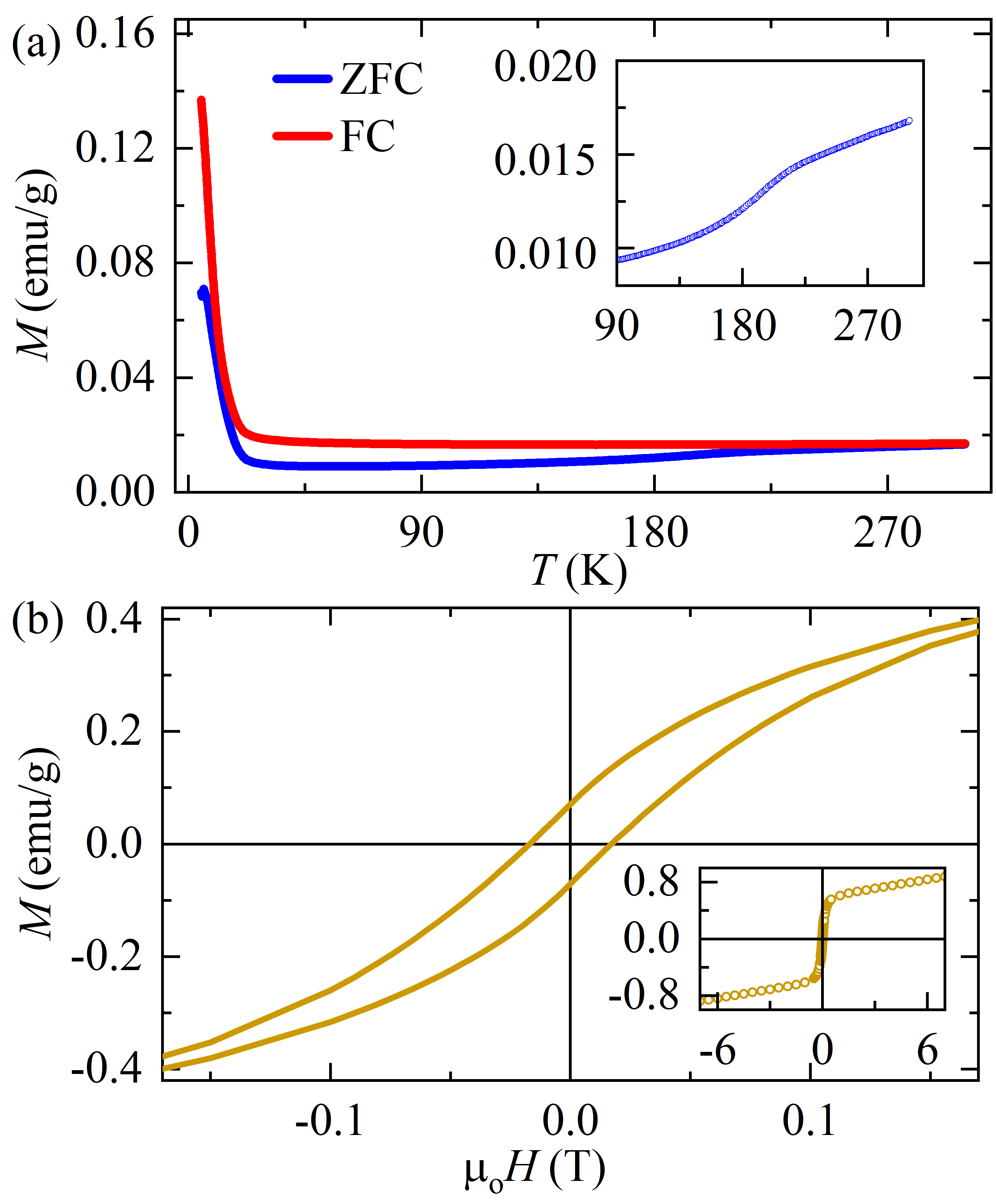}
	\caption{(a) Temperature-dependent magnetization (M) of (a) SiO$_2$-coated CoO8 nanoparticles measured in zero field-cooled (ZFC) and field-cooled (FC) protocols at an applied magnetic field ($\mu_oH$) of 100 Oe; inset shows the expanded view at high temperatures. (b) Magnetization versus field curve of SiO$_2$-coated CoO8 at 300 K.} 
	\label{fig9}
\end{figure*}

When we coated CoO8 with SiO$_2$, which is supposed to have the same crystallite size, it influences the magnetic response of nanoparticles quite interestingly. It should be mentioned here that although the thickness of the silica coating and hence the mass of the CoO/SiO$_2$ system is not precisely known, its thickness is believed to be smaller than the diameter of the nanoparticle~\cite{verma2017}. The temperature-dependent magnetization measured in ZFC and FC at 100 Oe is shown in Fig.\ref{fig9}(a). As temperature decreases, ZFC magnetization decreases slowly compared to CoO8 (Fig.\ref{fig4}(c)) and shows a steep increase below 20 K (Fig.~\ref{fig9}(a), inset). However, a clear hump associated with $T_N$ is seen near the same temperature of 213 K as found also in CoO8. This means SiO$_2$ coating just affects the surface moment but not the core of the nanoparticles. The magnetization measured in FC at 7 T is approximately independent of temperature down to 20 K, and then steeply increases as temperature further decreases, similar to ZFC magnetization.

The $M-H$ curve at 300 K of SiO$_2$-coated CoO nanoparticles is depicted in Fig.\ref{fig9}(b). It shows finite coercivity and remanence, although it does not saturate till $H = 7$ Tesla, which indicates a weak ferromagnetic response of CoO8 still present after SiO$_2$ coating. The value of the coercive field has been found to be around $H_C = 340$ Oe, which is significantly reduced compared to CoO8 (610 Oe). Note that both the $M-T$ and $M-H$ curves of SiO$_2$-coated CoO nanoparticles show distinct behavior compared to bare CoO8, which underlines the fact that SiO$_2$ coating affects the overall magnetic response of the CoO nanoparticles. SiO$_2$-coating on nanoparticles is expected to reduce agglomeration of nanoparticles and likely to reduce the interparticle interaction\cite{ambrose1996finite,larumbe2012}. Thus, the SiO$_2$-coating may affect the overall collective magnetic response. In addition, the magnetism at the interface between magnetic CoO nanoparticles and non-magnetic SiO$_2$ can be quite interesting and is a subject of detailed investigation~\cite{OjaPRL2012,hellman2017}.

Therefore, it is noteworthy to mention a few distinct changes in the magnetic response of Co/CoO nanocomposites and SiO$_2$-coated CoO8 nanoparticles compared to bare CoO8 nanoparticles:

(i) Co/CoO nanocomposites don't exhibit SEB like CoO8. The disordered surface spins on the surface of CoO8 possibly give rise to SEB, while ferromagnetic Co dominates over disordered surface spins, and hence SEB is not found in Co/CoO nanocomposites.

(ii) The magnetization of CoO8 is significantly reduced after SiO$_2$ coating. Since the magnetic moment in CoO8 mainly comes from surface spins, the reduction in the magnetic moment may indicate that surface spins are affected by SiO$_2$.

(iii) The $T\rm_N$ is observed near the same temperature of $\sim$ 213 K. This suggests that SiO$_2$ coating mainly affects the surface moment but not the core of nanoparticles.

(iv) The bifurcation between ZFC and FC curves has reduced. This indicates that spin disorder reduces after SiO$_2$ coating, possibly originating from the shell of nanoparticles.

(v) A steep rise in magnetization below 20 K, similar to the larger-sized sample CoO3, is observed. This is consistent with the reduction in magnetization with SiO$_2$, similar to an increase in size as in CoO3. 

The present study, therefore, suggests that both surface and finite-size effects play an important role in modifying the magnetic properties of these nanoparticles~\cite{ambrose1996finite,ambrose1996,larumbe2012}. As the crystallite size decreases, the finite-size and surface effects are clearly reflected: there is an increase in dangling bonds, local symmetry breaking, and site-specific surface anisotropy~\cite{Winkler2005}. Consequently, FM-like interactions are increased on the surface of the nanoparticles. These interactions reveal a core-shell magnetic structure, wherein each nanoparticle has an AFM ordered core and a disordered shell with uncompensated surface spins. The surface spins, in the present case, exhibit unconventional glassy behavior that is substantially different from the previously proposed spin-glass, superparamagnetic, and DAFF behavior~\cite{PhysRevLett.101.097206,salabacs2006exchange}. We attribute the SEB and CEB in these nanoparticles to the exchange coupling between the antiferromagnetically-ordered core and a ferromagnetically disordered shell that consists of uncompensated surface spins.

\section{Conclusion}  
We investigated the magnetic properties of CoO nanoparticles with sizes ranging from 10.8 nm to 34.6 nm. The significant role of the size effect is evident from the reduction in the Néel temperature with the decrease in crystallite size. The ac susceptibility of 10.8 nm nanoparticles shows frequency dependence, which is a characteristic of glass-like systems. The thermoremanent (TRM) and isothermoremanent (IRM) magnetization curves exhibit core-shell magnetic behavior; the core is antiferromagnetic, while the shell exhibits unconventional glassy behavior distinct from known spin-glass, superparamagnet, and diluted antiferromagnet in the field (DAFF) behaviors. Exchange coupling between the core and shell elucidates the observed spontaneous exchange bias (SEB) and conventional exchange bias (CEB) in 10.8 nm nanoparticles. This study paves the way for tuning exchange bias in single-phase nanoparticle systems through nanostructuring, which may be useful for identifying EB in other systems and their diverse range of applications.


\begin{acknowledgement}
	
Authors are grateful to Dr. N. P. Lalla and  Dr. D. Kumar, UGC-DAE Consortium for Scientific Research, Indore, India for providing TEM and synchrotron XRD data, respectively. The authors thank to USIF, Aligarh Muslim University, Aligarh for providing HRTEM data.  
		
\end{acknowledgement}


\providecommand{\latin}[1]{#1}
\makeatletter
\providecommand{\doi}
{\begingroup\let\do\@makeother\dospecials
	\catcode`\{=1 \catcode`\}=2 \doi@aux}
\providecommand{\doi@aux}[1]{\endgroup\texttt{#1}}
\makeatother
\providecommand*\mcitethebibliography{\thebibliography}
\csname @ifundefined\endcsname{endmcitethebibliography}
{\let\endmcitethebibliography\endthebibliography}{}



\end{document}